\documentclass[10pt,conference]{IEEEtran}
\IEEEoverridecommandlockouts
\usepackage{cite}
\usepackage{amsmath,amssymb,amsfonts}
\usepackage{algorithmic}
\usepackage{graphicx}
\usepackage{textcomp}
\usepackage{xcolor}
\usepackage{xspace}
\usepackage{comment}    
\usepackage{url}
\usepackage[normalem]{ulem} 
\usepackage{ifthen}
\usepackage{textcomp}
\usepackage{url}
\usepackage{array}
\usepackage{paralist}
\usepackage{fontawesome} 
\newcolumntype{C}[1]{>{\centering\arraybackslash}p{#1}}
\usepackage[inline, shortlabels]{enumitem} 
\usepackage[pdftex,colorlinks=true,pdfstartview=FitV,linkcolor=black,citecolor=black,urlcolor=black,breaklinks=true]{hyperref}

\newcommand{\ie}{\emph{i.e.},\xspace}
\newcommand{\eg}{\emph{e.g.},\xspace}
\newcommand{\etal}{\emph{et al.}\xspace}

\newcommand{\SO}{{SO}\xspace}
\newcommand{\ml}{{mailing lists}\xspace}

\newcommand{\figlabel}[1]{\label{fig:#1}}
\newcommand{\figref}[1]{\autoref{fig:#1}}

\newcommand{\rqI}{What do developers ask about commenting practices?}
\newcommand{\rqII}{What information do developers write in comments?}
\newcommand{\rqIII}{How do researchers support comment quality assessment?}

\def\BibTeX{{\rm B\kern-.05em{\sc i\kern-.025em b}\kern-.08em
    T\kern-.1667em\lower.7ex\hbox{E}\kern-.125emX}}
\begin{document}

\title{Speculative Analysis for Quality Assessment of Code Comments}

\author{\IEEEauthorblockN{Pooja Rani}
\IEEEauthorblockA{{Software Composition Group, University of Bern} \\
Bern, Switzerland
\\\href{http://scg.unibe.ch/staff/}{\faGlobe \hspace{0.1cm}scg.unibe.ch/staff}}
}

\maketitle

\begin{abstract}
Previous studies have shown that high-quality code comments assist developers in program comprehension and maintenance tasks.
However, the semi-structured nature of comments, unclear conventions for writing good comments, and the lack of quality assessment tools for all aspects of comments make their evaluation and maintenance a non-trivial problem.
To achieve high-quality comments, we need a deeper understanding of code comment characteristics and the practices developers follow.
In this thesis, we approach the problem of assessing comment quality from three different perspectives:
what developers ask about commenting practices, what they write in comments, and how researchers support them in assessing comment quality.

Our preliminary findings show that developers embed various kinds of information in class comments across programming languages. Still, they face problems in locating relevant guidelines to write consistent and informative comments, verifying the adherence of their comments to the guidelines, and evaluating the overall state of comment quality. 
To help developers and researchers in building comment quality assessment tools, we provide:
(i) an empirically validated taxonomy of comment convention-related questions from various community forums,
(ii)  an empirically validated taxonomy of comment information types from various programming languages,
(iii) a language-independent approach to automatically identify the information types, and
(iv) a comment quality taxonomy prepared from a systematic literature review.

\end{abstract}


\section{Introduction}

Well-documented code facilitates various software development and maintenance activities~\cite{Souz05a,Cioc96a}.
Several studies show that high quality code comments help developers in program comprehension~\cite{Deke09a}, suitable API selection~\cite{McMi10a}, and bug detection~\cite{Tan07c}.
However, comments are written using natural language sentences and their syntax and semantics are neither enforced by a programming language nor checked by the compiler.
As a result, developers are free to use numerous means and conventions to write comments~\cite{Padi09a}, and embed various types of information in them~\cite{Pasc17a}, thus making the quality evaluation of comments more complicated.

To guide developers in writing consistent and informative comments, programming language communities such as those for Java and Python, and large organizations such as Google and Oracle provide coding style guidelines.
However, these guidelines only marginally cover aspects of commenting code such as content, style, and syntax. 
Furthermore, the availability of several guidelines for a language makes developers unsure about
which comment conventions to use, which syntax to follow, and which type of information to write for what kinds of comments.
Therefore, developers ask questions on \ml, and community platforms such as Stack Overflow (\SO) and Quora to address these issues~\cite{Baru14a,Yang16a}.
Analyzing such developer concerns is valuable to understand their needs, and to identify challenges related to commenting practices.
Similarly analyzing their actual commenting practices is essential to understand the information they embed in comments and to ensure the quality of that information. 

Previous studies have characterized developer commenting practices in OOP languages by classifying comments based on the information that comments contain~\cite{Haou11a,Stei13b,Pasc17a,Zhan18a}.
Given the variety of comment types (class, method, or inline), not all comment types describe the source code at same levels of abstraction, therefore, the quality assessment tools need to be tailored accordingly. 
For example, class comments in Java should present high-level information about a class, whereas method comments should present implementation-level details\cite{Nurv03a}.
These commenting conventions vary across programming languages.
For instance, in comparison to Java, class comments in Smalltalk are expected to contain high-level design details and low-level implementation details.
Given the increasing usage of multi-language software systems~\cite{Toma14a} and persistent concerns about maintaining high documentation quality, it is critical to understand what developers write in a particular comment type, and to 
build tools to extract and check the embedded information across languages.
This can also help to ensure the extent to which a comment type adheres to a coding style guideline from the content aspect.

Even when a comment adheres to its coding style guidelines from all aspects such as content, syntax, and style, it is still possible that the comment is incomplete or inconsistent with the code, and thus lacks the desired high quality. 
Therefore, several other quality attributes that can affect comment quality need to be considered in the overall assessment of comments.
Researchers have proposed numerous comment quality evaluation models based on a number of metrics\cite{Kham10a,Padi09a} and classification approaches~\cite{Stei13b}.
However, a unifying comment quality taxonomy to express the purposes for which researchers evaluate comments, and which quality attributes they consider important and integrate frequently in their comment quality models or tools 
is still missing.

In summary, a good understanding of the existing practices that developers follow, and of comment quality models researchers suggest is necessary to bridge the gap between the notion of quality and its concrete implementation.
To gain this required understanding, we analyze code comments from various perspectives, of developers in terms of what they ask and what they write in comments, and of researchers in terms of what they suggest.
In this exploration of semantics embedded in the comments, we use \emph{speculative analysis}, by analogy with speculative execution (\eg branch prediction).
Previous studies have also shown how speculative analysis can be used to develop tools that inform developers early and precisely about potential consequences of their actions~\cite{Brun10d,Mucs13a}.
In our case, we are interested in supporting developers to ensure comment quality while writing or using comments for various development tasks.

The goal of this thesis is to investigate practices in code comment writing and evaluation in a stepwise manner to ultimately improve comment quality assessment techniques. I state my thesis as follows:

\vspace{2mm}
\noindent\fbox{%
	\parbox{\linewidth}{%
	\fontsize{9}{10}\selectfont
    \emph{
    Understanding the specfication of high-quality comments to build effective assessement tools requires a multi-perspective view of the comments.
    The view can be approached by analyzing (1) developer concerns about comments, (2) their commenting practices within IDEs, and (3) required quality attributes for their comments.
    }
    }%
}\\
\

This dissertation will focus on three main questions: 
\begin{itemize}
    \item  \emph{\rqI} Answering this can help in (i) identifying the key challenges developers face with current conventions and tools, and (ii) adapting their approaches accordingly.
    \item  \emph{\rqII} Understanding this can support the development of tools and approaches (i) to identify important information types and comment clones automatically, and 
    (ii) to verify the adherence of comments to style guidelines from the content aspect. 
    \item  \emph{\rqIII} Answering this question can help in (i) identifying the limitation of the existing tools and techniques, and (ii) adapting the tools according to the proposed comment quality taxonomy.
\end{itemize}

\section{Thesis Vision}

This section presents the studies conducted to answer the main questions, as shown in \figref{overview-dissertation}.
The following subsections briefly describe the motivation, methodology, and preliminary findings of each study.

\begin{figure}[ht]
    \centering
    \includegraphics[width=0.9\linewidth]{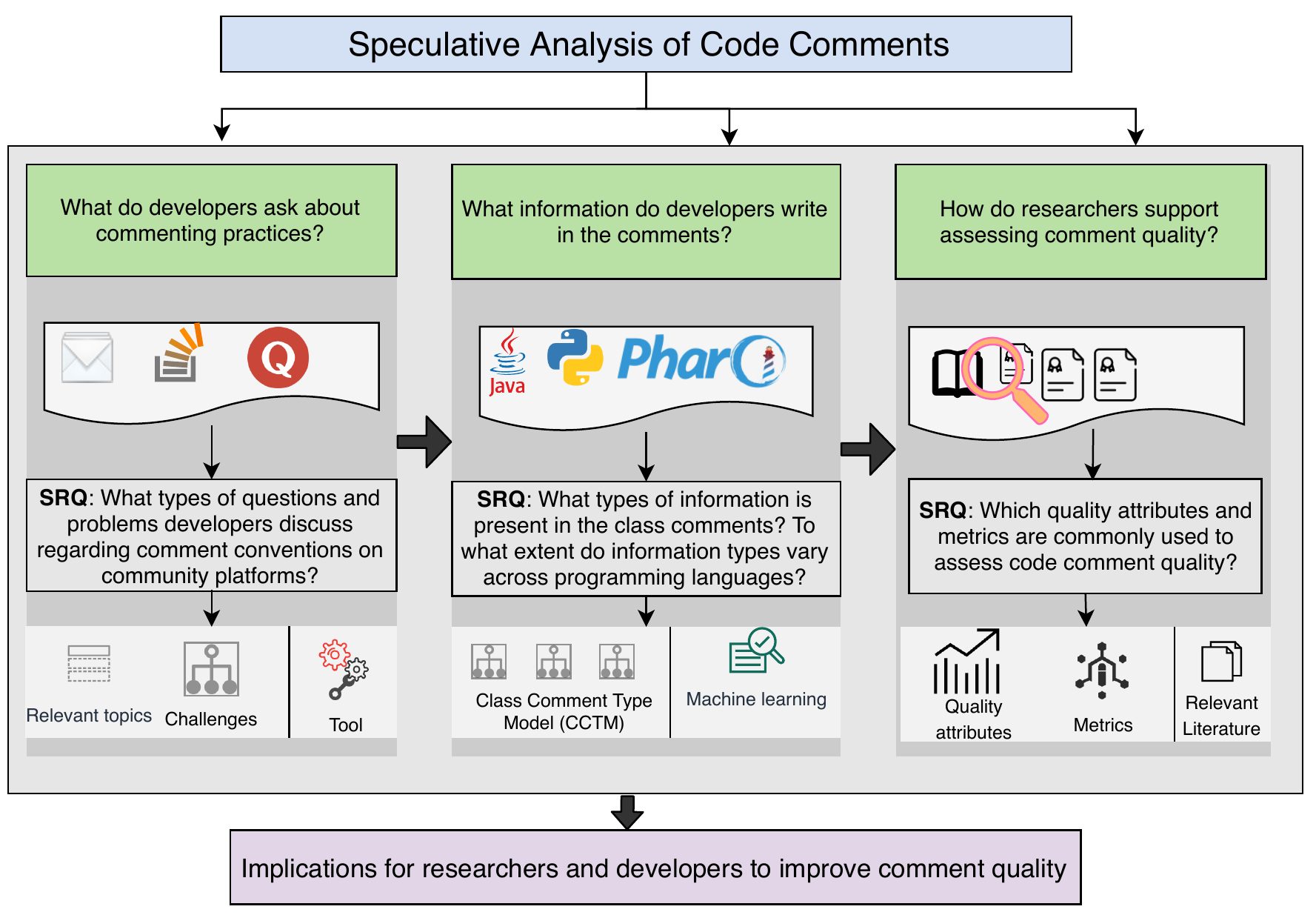}
    \vspace{-2mm}
    \caption{Overview of my dissertation with all research questions, their methodology, and results.}
    \figlabel{overview-dissertation}
    \vspace{-3mm}
    \end{figure}    

\subsection{\rqI}
\label{subsec:rqI}
Previous studies have leveraged various online platforms 
to gain a deep understanding of developers needs and challenges\cite{Baru14a,Yang16a,Agha19a}.
 We investigated the popular Q\&A forums, \SO and Quora, and Apache project-specific \ml to understand their commenting practices.

\textbf{Methodology}. To answer the question, we mined and pre-processed 11\,931 posts extracted using 14 relevant tags on \SO.
The tags are selected using a hybrid approach combining a heuristics-based approach used by Yang \etal ~\cite{Yang16a} and a keyword-based approach used by Aghajani \etal successfully in their work~\cite{Agha19a}.
We used a semi-automated approach based on Latent Dirichlet Allocation (LDA)~\cite{Blei03a},
 an advanced and popular topic modeling technique, to identify topics from the selected posts.
To uncover developer concerns in detail (mainly which type of questions they ask and about what tools and techniques), we manually analyzed a statistically significant sample set of posts from \SO, Quora and \ml, and formulated a taxonomy of these concerns.
The taxonomy offers an overview of the leading questions discussing commenting conventions in a more formal, structured, and possibly exhaustive way.

\textbf{Findings}. Our study results highlight that: 
\begin{enumerate*}[(i)]
    \item Developers ask questions about best practices to write comments (15\% of the questions) and generate comments automatically using various tools and technologies. 
    \item Among 14 topics identified by LDA, we found five irrelevant topics due to the generality and commonality of the tags (\eg ``convention,'', ``commenting'').
    \item  From our manual analysis, we found that developers are interested in embedding various kinds of information, such as code examples and media (\eg images) in their code comments but lack clear guidelines to write them. 
    \item Developers post questions about documentation tools on \SO, whereas no such questions are reported on Quora. In \ml, we did not find enough developer discussions about comment conventions.
\end{enumerate*}\\
\textbf{Conclusion}.
This analysis shows that developers use various community platforms to raise concerns about code comments.
Such concerns hint at the challenges developers face, and their needs from the programming language communities, tools, technologies, and researchers.
Conveying clear guidelines to write good comments,
and building tools to verify the adherence of comments to these guidelines
 indicate possible directions to support developers.

\subsection{\rqII}
\label{subsec:rqII}
Source code comments consist of several comment types (class comments, method comments, inline comments), but not all comment types contain the same types of information.
We start our analysis by first focusing on class comments, which play an important role in obtaining a high-level overview of classes in object-oriented programming languages~\cite{Clin15a}.
Class commenting practices however vary across programming languages.
For instance, a class comment in Java or Python contains high-level overview details and uses annotations (\eg @param,@return) to express specific types of information.
In contrast, class comments in Smalltalk contain detailed design and implementation documentation, and they do not make use of any annotation.
We first investigated class comments in Pharo (a modern Smalltalk environment), and identified the types of information developers embed in them by studying the research question \textbf{RQ$_1$}: \emph{What types of information are present in Pharo class comments?}
Then we measured the adherence of Pharo class comments to the class comment guidelines.
To generalize our findings across languages, we extended our analysis to other programming languages, namely Java and Python.
We systematically compared the commonalities and differences among class commenting practices with the research question \textbf{RQ$_2$}: \emph{To what extent do information types vary across programming languages?}
In order to automate the identification of information types from class comments across languages, we studied the research question \textbf{RQ$_3$}: \emph{Can machine learning be used to identify class comment types according to our taxonomy automatically?}

\textbf{Methodology}. To answer {RQ$_1$}, we conducted a three-iteration-based analysis on a {statistically significant} sample set of 714 comments selected from internal and external projects of Pharo.
Three authors analyzed the content of comments using open-card sorting and pair sorting 
to build and validate the comment taxonomy.
In the case of Python and Java class comments, we used the initial comment taxonomy available from previous works~\cite{Pasc17a,Zhan18a}, and analyzed and validated the content using the closed-card sorting technique.
Based on the constructed taxonomy \ie Class Comment Type Model (CCTM) and labelled data from each language, we answered {RQ$_2$}.
To automatically classify class comment types according to CCTM for {RQ$_3$}, we used an approach that leverages two techniques — namely Natural Language Processing (NLP) and TF-IDF.
We use the TF-IDF technique as a baseline due to its successful adoption in recent work on classifying code comments~\cite{Misr20a}.
We transform a multi-label classification into a set of single-label classification problems to balance one label at a time and avoid over-fitting the categories.
We adopt a 10-fold cross-validation strategy with a standard probabilistic Naive Bayes classifier, the J48 tree model, and the Random Forest model based on the recent work~\cite{Pasc17a}.
We evaluate {RQ$_3$} by measuring precision, recall, and F-measure of our approach against the TF-IDF baseline approach.

\textbf{Findings}.
Our results highlight that:
\begin{enumerate*}[(i)]
    \item Developers express different kinds of information (more than 15 information types) in class comments ranging from the high-level overview of the class to low-level implementation details across programming languages.
    \item Class comments contain various types of information but not all of these information types are suggested by coding guidelines, and this behaviour is observed across all the selected languages.
    In the case of Pharo, the information types suggested by the guidelines were observed more frequently than other information types. 
    We are in still in the process of verifying this observation in other programming languages.
    \item The Random Forest algorithm fed by the combination of NLP+TF-IDF features achieves the best classification performance for the top six frequent categories over the investigated languages with relatively high precision (ranging from 78\% to 92\% for the selected languages), recall (ranging from 86\% to 92\%), and F-Measure (ranging from 77\% to 92\%) where Pharo achieves less stable results compared to Python and Java.
\end{enumerate*}

\textbf{Conclusion.}
This analysis highlights the diverse types of information developers embed in class comments regardless of coding style guidelines about comments. Given the benefits of retrieving these information types automatically for various development tasks, it highlights the challenges in unifying retrieval approaches across languages.

\subsection{\rqIII}
\label{subsec:rqIII}
Software quality is frequently represented as a contextual concept.
Therefore,  it requires identification and quantification of important characteristics of high-quality software as a first step to measure it~\cite{Bans02a}.
The main objective of our literature review is: to identify the quality attributes that are used to assess code comment quality and collect the metrics used to measure these quality attributes.
Additionally, we are interested in which tools/models have been proposed by researchers to assess comment quality.
To achieve these objectives, we plan to conduct a systematic literature review (SLR) to answer the following research questions.
\textbf{RQ$_1$}: \emph{What quality attributes are used to evaluate the quality of code comments, and what metrics are used to estimate the quality attributes?}
\textbf{RQ$_2$}: \emph{Which quality attributes and metrics do current assessment tools support?}

\textbf{Methodology}: We plan to conduct the SLR following the guidelines of Kitchenham \cite{Kitc07a}.
We separate the study steps associated with the SLR-related phases planning, conducting the review, and reporting. 
In the planning phase, we identify the objectives of our SLR and specify the research questions. 
We plan to review the proceedings of the past ten years \ie 2010-2020 from the relevant SE conferences and journals according to the \emph{Computing Research and Education Association of Australasia} (CORE) ranking.\footnote{\emph{CORE rankings portal}, accessed August 18, 2020, \url{http://www.core.edu.au/}}
We formulate the inclusion and exclusion criteria. 
Based on these criteria, we plan to systematically identify relevant studies.

\textbf{Expected output}. Insights from the SLR are expected to provide a detailed view of the tools and techniques proposed by researchers to assess the quality of comments. 
Based on these insights, we plan to prepare a comment quality taxonomy which can help researchers and developers in identifying various quality attributes suitable for a comment type and integrating the relevant measures in their tools as per their requirements.

\section{Preliminary and Expected Contributions}
From each study, we present empirical insights, approaches, and tools to support developers and researchers in ensuring high-quality comments.

\begin{itemize}
    \item For the first question \emph{``\rqI''}, we present:
\begin{inparaenum}[(i)]
  \item an empirically validated taxonomy of comment convention-related questions from various community forums, and
  \item a tool to conduct a mining study on multiple sources or forums,
\end{inparaenum}

\item For the second question \emph{``\rqII''}, we provide:
\begin{inparaenum}[(i)]
  \item an overview of the Pharo class commenting trends over seven major releases till 2019,
  \item an empirically validated taxonomy, called CCTM, characterizing the information types found in class comments written by developers in three different programming languages, and
  \item an automated approach (available for research purposes) able to accurately classify class comments according to CCTM
\end{inparaenum}
\item 
For the third question \emph{``\rqIII''}, we expect to achieve: 
\begin{inparaenum}[(i)]
  \item a comment quality taxonomy to identify relevant quality attributes, and
  \item a review of existing tools and techniques that assess the quality of code comments.
\end{inparaenum}
\end{itemize}

\section{Proposed Timeline}
I am a third-year PhD student and will be entering the final year of my PhD from January 2021.
The expected timeline for the projects:
\begin{itemize}
    \item  A first study (\ref{subsec:rqI}) has been submitted to the journal-first track at Transactions on Software Engineering and Methodology, 2020 (TOSEM'20).
    \item The RQ1 in the second study 
    (\ref{subsec:rqII}) is currently undergoing a minor revision in the journal-first track at Empirical Software Engineering (EMSE'19)\cite{Rani20a}.
    \item Other research questions (RQ2 and RQ3) in the second study (\ref{subsec:rqII})
     have been submitted to the Journal of Systems and Software (JSS'20).
    \item The third study (\ref{subsec:rqIII}) is currently planned to be submitted to Transactions on Software Engineering (TSE'21).
\end{itemize}

\section{Related work}
\textbf{Comment conventions (RQ1)}:
Developers frequently use various web resources to satisfy their information needs.
Recently, researchers have started leveraging these resources such as version control systems~\cite{Chen16a}, 
Q\&A forums~\cite{Yang16a,Baru14a}, and \ml~\cite{Agha19a}.
In the context of software documentation, Aghajani \etal studied documentation issues on \SO, Github and \ml~\cite{Agha19a} and formulated a taxonomy of these issues. 
However, they have focused on the issues related to project documentation, such as wikis, user manuals, and code documentation, and do not focus specifically on the issues of the convention of the code comments.
Barua \etal found questions concerning coding style and practice to be amongst those most frequently appearing on \SO~\cite{Baru14a}, but did not investigate it further.
Our first question (\ref{subsec:rqI}) focuses specifically on the problems related to commenting practices developers discuss on \SO, Quora, and \ml.

\textbf{Identify information types from comments (RQ2)}:
Code comments contain valuable information to help developers in various activities and tasks.
Pascarella \etal identified the information types from Java code comments and presented a taxonomy~\cite{Pasc17a}.
Similarly, Zhang \etal identified information types from Python code comments~\cite{Zhan18a}.
We focused specifically on class commenting practices and used their taxonomy as an initial taxonomy to classify class comments. 
Compared to the work of Pascarella \etal and Zhang \etal~\cite{Pasc17a,Zhan18a}, we found several other types of information such as warnings, observations, and recommendations developers embed in class comments. 
To identify different kinds of information from comments automatically, several studies have explored numerous approaches based on heuristics or textual features~\cite{Drag10a,Shin18a}.
In contrast to these previous approaches, we extracted the natural language patterns (heuristics) automatically using a tool, combined them with other textual features, and tested our approach across languages. 

\textbf{Comments quality (RQ3)}:
Apart from identifying information embedded in the comments, assessing comments from other perspectives has gained a lot of attention from researchers in the past years,
for example, 
assessing comment quality~\cite{Kham10a,Stei13b}, detecting inconsistency between code and comments~\cite{Rato17a,Wen19a}, and examining co-evolution of code and comments~\cite{Flur09a}.
The main aim is to keep comments consistent with the code and to maintain their high quality.
Several recent works have proposed tools and techniques to automatically assess the comments using specific quality attributes and metrics~\cite{Kham10a,Stei13b,Yu16a}.
However, a unifying model of comment quality attributes and metrics that are considered important for assessing comments is still missing. 
Previous literature reviews have provided the quality models for the software documentation~\cite{Ding14a,Zhi15a} but we focus specifically on the code comment aspect.

\section{Conclusion}
To improve the state of comment quality assessment techniques, this thesis focuses on three main questions: what do developers ask about commenting practices, what do they write in comments, and how do researchers support assessment of comments.
Our work draws insights from both empirical evidence mined from developer sources and research results (SLR).
Our preliminary findings show that developers embed various kinds of information in comments. 
Still, they face several problems in locating the specific comment guidelines, verifying the adherence of their comments to the coding standards, and evaluating the overall state of the comment quality.
Our empirical evidence also shows that Pharo developers follow commenting guidelines in writing class comments.
These insights of developer commenting practices across languages can help researchers to improve comment quality assessment tools, and to evaluate comment summarization and comment generation approaches.
We present initial approaches, tools, and labelled dataset to facilitate the future comment analysis work on other languages and environment.

My future work will concentrate on exploring which tasks and activities require developers to search these information types in comments and how developers find these information types. 
Based on developer commenting practices, my objective would be to improve my prototype tools and reduce the efforts in assessing comment quality.
I expect to finish the work for my dissertation in 2021.

\bibliographystyle{IEEEtran}
\bibliography{scg}

\end{document}